\documentclass[11pt,a4paper]{article}
\usepackage{jheppub}

\usepackage{amssymb}
\usepackage{epsfig}
\newcommand{\be}{\begin{equation}}
\newcommand{\ee}{\end{equation}}
\newcommand{\ba}{\begin{eqnarray}}
\newcommand{\ea}{\end{eqnarray}}
\newcommand{\bd}{\begin{displaymath}}
\newcommand{\ed}{\end{displaymath}}
\newcommand{\al}{\alpha}
\newcommand{\bt}{\beta}
\newcommand{\ga}{\gamma}

\newcommand{\da}{\delta}

\newcommand{\za}{\zeta}

\newcommand{\Da}{\Delta}

\newcommand{\cF}{{\cal F}}
\newcommand{\cP}{{\cal P}}

\newcommand{\cO}{{\cal O}}

\newcommand{\p}{\partial}

\newcommand{\ra}{\rightarrow}
\newcommand{\Ra}{\Rightarrow}

\newcommand{\LF}{\left(}
\newcommand{\RF}{\right)}
\newcommand{\LT}{\left[}
\newcommand{\RT}{\right]}

\newcommand{\mx}{\mbox}
\newcommand{\mt}{\mathtt}
\newcommand{\mand}{\mx{ and }}

\newcommand{\with}{\mx{ with }}

\newcommand{\vs}{\vspace{1mm}\\}

\begin{document}

\title{Atick-Witten Hagedorn Conjecture, near scale-invariant matter and blue-tilted  gravity power spectrum}

\author{Tirthabir Biswas$^1$}

\author{Tomi Koivisto$^{2}$}

\author{Anupam Mazumdar$^3$}

\affiliation{
{\it $^1$Department of Physics,
Loyola University, New Orleans, LA 70118}\\
{\it $^2$Nordita, KTH Royal Institute of Technology and Stockholm University,
Roslagstullsbacken 23, SE-10691 Stockholm, Sweden}\\
{\it $^3$Consortium for Fundamental Physics, Physics Department, Lancaster University, LA1 4YB, UK}
}

\date{\today}

\abstract{
We will provide an interesting new mechanism to generate almost scale invariant seed density perturbations with a {\it red spectrum},
while keeping the gravitational wave spectrum  {\it blue-tilted} in a stringy thermal contracting phase at temperatures beyond the Hagedorn temperature. This phase is often referred to as
the Hagedorn phase where the free energy has been conjectured by Atick and Witten to grow more slowly than ordinary radiation. The primordial fluctuations are created by the statistical thermal fluctuations determined by the partition function, rather than quantum vacuum driven fluid dynamical  fluctuations. Our mechanism assumes a non-singular bounce.}

\preprint{NORDITA-2014-37}

\maketitle

\noindent

There has been a very important discovery of B-mode of polarisation from the BICEP team~\cite{Ade:2014xna}, which points  its origin towards the
primordial gravitational waves seeded during the earliest epochs in the history of the universe, roughly at the $10^{16}$~GeV scale.
The data suggests a tensor-to-scalar ratio
\be
 0.15\leq  r(k_\ast) \equiv {\cal P}_{\zeta}(k_\ast)/{\cal P}_{h}(k_\ast) \leq  0.27\ ,
\ee
at the pivot scale,
$k_{\ast} = 0.002~{\rm Mpc}^{-1}$. There is a mild hint that the gravitational wave spectrum is growing with a {\it blue-tilted} spectrum~\cite{tilt-analysis1,tilt-analysis2}, while
Planck~\cite{Planck-1} and WMAP~\cite{Hinshaw:2012aka} has provided evidence for a {red tilted} matter power spectrum. Typically, in inflationary models it is challenging to
generate a {\it blue tilted} gravitational wave. The amplitude of the gravitational waves is proportional to the total energy density of the universe, the energy density needs to grow as progressively shorter modes  leave the Hubble patch during inflation. However, this is impossible in a monotonically  expanding universe as long as the matter content satisfies the weak energy condition. For instance, the Hubble expansion rate decreases very gradually during typical inflation, for a review, see~\cite{Mazumdar:2010sa}. Now, in bouncing (cyclic) cosmologies there is a (are) {\it contracting phase(s)} where the energy density increases preceding the {\it expanding phase(s)}. This therefore naturally gives rise to a {\it blue-tilted } gravitational wave
spectrum. The challenging part, actually, is  to produce a near-scale-invariant red-tilted scalar spectrum.

In this paper we will first present a new mechanism to generate a near scale-invariant spectrum of scalar perturbations based on a thermal stringy Hagedorn phase. We will see that the tilt of the scalar spectrum can be either red or blue depending upon the parameters of the model and therefore can be consistent with Planck and WMAP observations. Next, we will calculated the gravity wave spectrum in this model. We will see that the model comes with a definite prediction of a blue-tilted spectrum\footnote{In the context of string cosmology a blue tilt of relic gravitons may be generic, as already pointed out in Ref. \cite{Brustein:1995ah}.}, and that it can give rise to a relatively large tensor-to-scalar ratio consistent, in particular, with the observed value by the BICEP team.

One of the most intriguing features of string theory is the existence of the Hagedorn phase at high temperatures where the energy is not dominated by the massless modes, but rather by the most {\it massive string states}, leading to a pressureless fluid~\cite{jain,vafa,robert}. In fact, a canonical description of the thermal phase indicated a limiting Hagedorn temperature~\cite{jain}.  Later, however, it was argued that the limiting temperature only corresponds to the emergence of a thermal tachyonic mode making the description of the system in terms of fundamental string excitations invalid~\cite{bala-old}. It was further conjectured by Atick and Witten in one of the classic papers~\cite{atick}  that at temperatures larger than the Hagedorn temperature, the free energy $\cF$ grows much more slowly,
\be\label{free-0}
\cF\propto T^2\ ,
\ee
as compared to conventional field theories where $\cF\propto T^4$. Therefore, the system represents many fewer degrees of freedom than one would have expected from the zero-temperature string spectrum, or even in point-like particle field theories. It is worth pointing out that a finite temperature loop calculation~\cite{padic} of a toy $p$-adic string model precisely exhibited the $T^2$ behavior along with thermal duality: $Z(T)\sim Z(T_H^2/T)$, another feature also conjectured in~\cite{atick}. $Z$ denotes the finite temperature partition function and $T_H$, the Hagedorn temperature, is related to the string scale via $\cO(1)$ factors. Finally, we note that such a ``stiff fluid'' (equation of state parameter, $w=p/\rho\approx 1$) behaviour (\ref{free-0}) has also been argued to emerge in a quantum gravity phase of interconnected blackholes and holographic principles~\cite{Banks-fischler}. Surprisingly, (\ref{free-0}) seems also to  lead to a new mechanism to generate near scale-invariant matter perturbations.

We should point out that our scenario is different from the Brandenberger-Nayeri-Vafa mechanism~\cite{Brandenberger}, which is  based on the behaviour of closed string modes below the Hagedorn temperature. In our case  we are looking at stringy thermodynamics above the Hagedorn temperature. Moreover, the mechanism in \cite{Brandenberger} requires a loitering or a slow bounce phase~\cite{hagedorn-bounce} to realize which one has to invoke new physics beyond  Einstein's  theory of General Relativity (GR), whereas the mechanism we will discuss is based on a contracting universe dominated by a stiff fluid and evolving according to GR. We  require a bounce mechanism (see e.g. \cite{nonsingular,fermion-condensation})
only to transition from contraction to expansion, the modes that we are observing today at CMB exits prior to the bounce.
Our scenario is also very distinct from the ekpyrotic case~\cite{ekpyrotic} which operates at $w\gg 1$ whereas we have $w\approx 1$. More importantly, we are considering thermal fluctuations as opposed to scalar field fluctuations.
\section{Thermal Fluctuations}
The possibility of thermal fluctuations being the origin of small inhomogeneities and anisotropies in the cosmic microwave background perhaps dates back to Peebles~\cite{Peebles:1994xt}. In general, fluid fluctuations can arise from two different sources:
\begin{itemize}
\item There could be fluctuations in the energy density and the associated temperature. This can arise, for instance, due to quantum vacuum fluctuations as has been widely studied, see~\cite{BFM}.

\item However, even if one can define a unique temperature in a given volume, there are fluctuations in energy within this volume due to the very statistical nature of thermal physics. This could also potentially seed primordial fluctuations, see for instance~\cite{thermal} and references there-in.
\end{itemize}
The statistical fluctuations in the energy inside a given volume $L^3$ is given by
\ba
<\Da E>_L^2&\equiv& <E^2>-<E>^2={\p^2 \ln Z\over \p \bt^2}=T^2 C_L\nonumber\\
\Ra <\da \rho^2>_L&=& {T^2 C_V\over L^6}= {T^2\over L^{3}}{ \p \rho\over   \p T}
\ea
where  $C_L$ is the heat capacity of the thermal system for a given volume $L^3$.  These are random fluctuations that exists in any finite temperature system and arise already at the classical level as long as the fluid is in local thermal equilibrium.
The power spectrum for the seed perturbations could then be sourced by these thermal fluctuations till the wavelengths of the fluctuations are smaller than the Hubble length.  Once the modes become super-Hubble, thermal correlations over the relevant physical wavelengths can no longer be maintained,  the coupled  metric and matter fluctuations now evolve according to the usual General Relativistic hydrodynamical differential equations. Essentially, in this is a set-up  the sub-Hubble thermal fluctuations (instead of the traditional quantum vacuum fluctuations) act as the initial conditions for the
super-Hubble fluctuations.

Very recently in \cite{Biswas:2013lna} a precise understanding of how these statistical fluctuations get encoded in the curvature perturbations, $\za$, at the ``Hubble crossing'', was achieved  for a general extensive thermodynamic fluid whose pressure, $p$, can be an arbitrary function of the temperature, $T$. The derived curvature power spectrum reads
\ba
\cP_{\za}=k^3 \zeta_k^2  =  \sqrt{3}\gamma^2A^2(T_k)\frac{T_k^2\rho_k'}{M_p^3\sqrt{\rho_k}}\,,\label{spectrum}
\ea
where $\ga=2\sqrt{2}\pi^{3/4}\approx 6.7$ and $M_p$ is the reduced Planck mass.
The subscript $k$ (which we are going to subsequently drop) refers to the fact that all these quantities have to be evaluated at the Hubble crossing condition, $H_k=k/a$. We are using the notations described in~\cite{Biswas:2013lna}
\ba
&\mx{where }&\ \ \ \ \ \ A(T)=\frac{3(1+w)\Omega + 2(3+r)}{6(1+w)\Omega}\\
 \label{rate2}
&\mand&\ \ \ \ \ \ \ r=-{3\over 2}\LT1+\frac{\LF 1+w\RF\rho\LF 2\rho'+T\rho''\RF}{T{\rho'}^2}\,\RT.
\ea
Note, that all the above functions of temperature can be calculated if we know $p(T)$ as the energy density is related rather straightforwardly to pressure:
\be
\rho(T)=T{d p(T)\over d T}-p(T)
\ee
Just to illustrate, for radiation the above formula yields
\be \label{r_spectrum}
\cP_{\zeta}  = \frac{\sqrt{3 g}\ga^2}{4}\LF\frac{T}{M_p}\RF^3\,.
\ee
$g$ is the number of degrees of massless modes. $T$ corresponds to the temperature when the given mode becomes super-Hubble. Evidently, the spectrum depends strongly on the temperature leading to a large blue tilt~\cite{Biswas:2013lna}.
\section{Hagedorn Phase and CMB Spectrum}
According to the Atick-Witten conjecture, the partition function only grows as $T^2$ at high temperatures, it is then natural to assume the pressure to be of the form
\be\label{con-0}
p(T)=M_s^4\LT\LF {T\over M_s}\RF^2+c_1\LF {T\over M_s}\RF +c_2\ln\LF{T\over M_s}\RF\RT\,,
\ee
with subleading linear and log terms.
The  Atick-Witten behaviour should hold above $T_H$ which should be close to $M_s$, as both the scales are expected to be related to the string tension scale by numerical factors~\cite{atick,padic}.
 It is a little more transparent and convenient to work with a slightly different functional form
\be\label{con-1}
p(T)=M_s^4\LT\LF {T\over M_s}\RF^2+c_1\LF {T\over M_s}\RF^{1+\al}\RT\with |\al|\ll 1\,.
\ee
Hence, here we will focus on Eq.~(\ref{con-1}) while the analysis with Eq.~(\ref{con-0}) and more general subleading corrections will be provided elsewhere.

The spectrum can be calculated very straightforwardly, and one obtains:
\ba
\cP_{\za} &=& \frac{\sqrt{3}\gamma^2}{32} \left(1+\al - 3\al^2+\al^3\right)^2 c_1^2 \LF{M_s\over M_p}\RF^3 \LF{T\over M_s}\RF^{2 \alpha} \nonumber \\
&+&\cO\LF{T\over M_s}\RF^{-1+\al} \approx \frac{\sqrt{3}\pi^{\frac{3}{2}}c_1^2}{4} \LF{M_s\over M_p}\RF^3 \LF{T\over M_s}\RF^{2 \alpha} \label{Pzeta}
\ea
As $\al\ra 0$ we have a scale-invariant spectrum. If $\al>0$ the spectrum has a blue tilt, but if $\al<0$, the spectrum will be red-tilted. Since for a stiff fluid, $\rho\sim H^2\sim T^2\sim a^{-6}$,
the spectral tilt can be calculated quite easily:
\be
1-n_s=-3\al
\ee
If $c_1\sim \cO(1)$, to reproduce the correct amplitude of the power spectrum we should have $M_s/M_p\sim \cO(10^{-3}-10^{-4})$. These numbers are very reasonable and as one can see there is no fine-tuning required to generate the near scale invariance of the spectrum. One only requires the subleading correction to be close to being linear, a very natural assumption. In our opinion, this is a real advantage of our mechanism over standard inflationary models as it bypasses the need to have a very flat potential with small slow-roll parameters!
\section{Gravitational Waves}
The initial conditions for the primordial gravitational waves are set by quantum vacuum fluctuations since  extensive thermal matter does not provide any additional source for gravitational waves.  The gravitational wave spectrum is thus given by~\cite{Biswas:2013lna}
\be
\mathcal{P}_h = \frac{1}{4\pi^2}\LF \frac{H}{M_p}\RF^2 = \frac{\rho}{12\pi^2 M_p^4}\,.
\ee
The tensor to scalar ratio becomes
\ba
r &\equiv& {\cP_h\over \cP_{\za}}  \approx \frac{1}{3\pi^3\sqrt{3\pi} c_1^2}\LF{M_s\over M_p}\RF\LF{T\over M_s}\RF^{2(1- \alpha)}\,.
\label{rst}
\ea
It is clear that there is a large parameter space where $r$ is going to be appreciable. For instance, if $c_1\sim 0.1$, then in the region $T\sim \sqrt{M_sM_p}$ we will get $r\sim \cO(1)$.

\begin{figure}[htbp]
\begin{center}
\includegraphics[width=0.5\textwidth,angle=0]{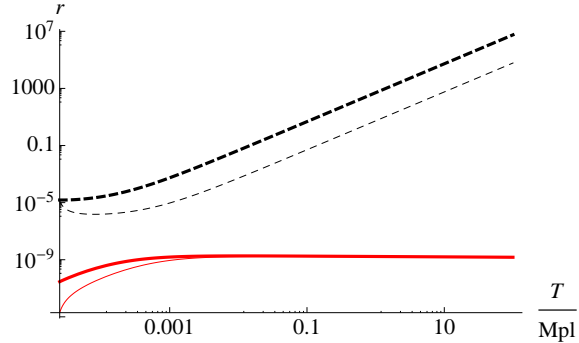}
\end{center}
\caption{ The scalar spectrum $\cP_{\za}$ (red solid line) and $r$ as (black dashed line) as functions of the temperature $T/M_p$. 
We have set $\al=-0.075$. For the thick lines $M=10^{-2}M_p$, and $c_1 \approx 0.9$. For the thin lines,  $M=10^{-3}M_p$, and $c_1 \approx 0.03$.
 \label{rTS1}}
\end{figure}

\begin{figure}[htbp]
\begin{center}
\includegraphics[width=0.5\textwidth,angle=0]{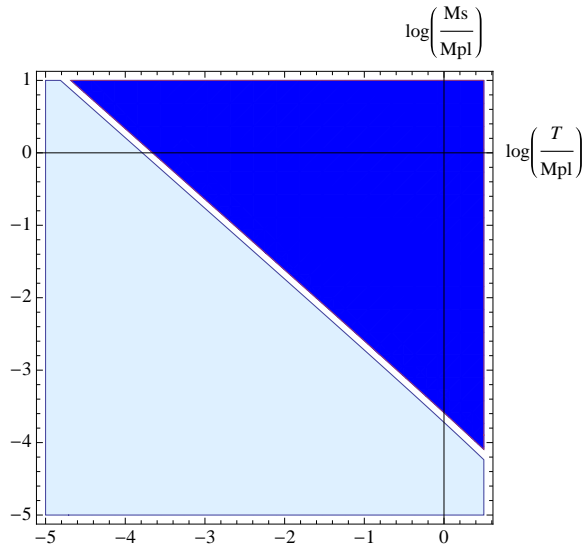}
\end{center}
\caption{The region of the parameter space where $0.15<r<0.27$ corresponding constraints from BICEP2~\cite{Ade:2014xna} is the patch in the middle. In the dark blue region we have $r>0.27$, and in the light region below $r<0.15$. The observable CMB scales span about 4 orders of magnitude on the vertical temperature axis for a given $M_s$.
  \label{rTS2}}
\end{figure}

Let us now consider the predictions in some details. To match with the observed CMB amplitude~\cite{Planck-1}, $A_0 \approx 2.4\cdot 10^{-9}$, we fix the parameter $c_1$ as
\be \label{c1}
c_1 = \frac{2\sqrt{A_0}}{3^\frac{1}{4}\pi^\frac{3}{4}}\LF{M_s\over M_p}\RF^{-\frac{3}{2}}\LF{T_{pivot}\over M_p}\RF^{-\al}\,.
\ee
In the following, we will use $T_{pivot}=0.01M_p$. In Fig.~\ref{rTS1}, we show the scalar spectrum and $r$ as functions of temperature when $M_s=10^{-4}M_p$~\footnote{The observed scales in CMB sky approximately span three orders of magnitude. Minimally we want all these modes to exit during the Hagedorn contraction phase, and since $T_k\propto k^{3/2}$, this corresponds to a temperature range spanning approximately $\cO(10^{4}-10^5)$. Since we expect (\ref{con-1}) to remain valid within $ T_H\sim M_s\lesssim T\lesssim \cO(10)M_p\sim$ Planck mass, there is a separate good reason to choose $M_s/M_p\sim \cO(10^{-3}-10^{-4})$.}. The value of $c_1$ is adjusted to match with the CMB power-spectrum according to (\ref{c1}), and we have used $\al=-4/300$ to obtain the central value of the spectral index, $n_s=0.96$, as observed by the PLANCK collaboration~\cite{Planck-1}. The interesting regime is $T \gtrsim M_s$, much below this, the spectrum is not scale-invariant, gravitational waves are unobservably small, and in any case we expect (\ref{con-1}) to be  valid only when $T\gtrsim M_s$. In Fig.~\ref{rTS2}, we have provided numerical plots of $r$ as a function of the ratios $T/M_p$ and $M_s/M_p$. The value of $c_1$ is adjusted to match with the CMB power-spectrum. As expected, we find large parameter spaces where $r$ can be considerable.
\section{Spectral tilts \& Low Multipole anomalies}
It is clear from the plots that the gravity waves have a strong blue tilt in our model which discriminates it from inflationary ones. In fact, it is easy to calculate the spectral tilt: Since $\mathcal{P}_h\sim T^2$
\be
n_t={d\ln \cP_h\over d\ln k}\approx 3
\ee
which is consistent with the value, $n_t= 1.5261^{+3.4739}_{-3.5261}$, that was   quoted  in a recent analysis~\cite{tilt-analysis1} performed by combining the BICEP2 data with Planck, WMAP, and BAO, see also~\cite{tilt-analysis2}. We would like to point out however, an important twist that a Planck scale bounce, that is anyway a requirement of the model,  may provide. While in the GR contracting phase $|H|$ increases, as the temperatures approach the Planckian regime, the increase must taper down (and eventually $H$ should start to decrease to reach $H=0$  at the bounce point). Since $\cP_h\propto H^2$, this would essentially decrease the observed $|n_t|$ (or give rise to a negative running of the tilt), the details of which will depend on  the nature of the bouncing background cosmology and requires further study. We note in passing that no assumption about the validity of GR is made while deriving the thermal power-spectrum (\ref{spectrum}) and it should  therefore remain valid near the bounce.

Finally, there is one last interesting possibility worth mentioning. Once $T\lesssim T_H$, we do not expect (\ref{con-1}) to remain valid, but rather to see a sharper increase of pressure with temperature. This would  change the dependence of the matter spectrum on temperature. In fact, for any $p\propto T^{2+\bt}$ with $\bt>0$, one obtains a blue spectrum. In other words, at around $T\sim M_s\sim T_H$, we expect to see a transition of the matter spectrum from a blue to a red tilt! It is tempting to investigate whether this feature can provide an explanation for the low multipole anomalies, the fact that the matter spectrum seems to rise between $\ell=2$ to $\ell\sim 40$~\cite{Planck-1}.

\section{Inflationary Stretching \& Closing Remarks}

Last, but not the least, it is important to highlight that these statistical fluctuations in the Hagedorn phase  need to be ``stretched'' if they are to seed the CMBR anisotropies and large scale structure formation. Although seldom discussed in any details, this is  an issue that has to be addressed in any cyclic/bouncing scenario, not just in the scenario we have proposed.  The issue is the following: since the perturbations are expected to  ``freeze'' after they exit the Hubble radius, the amplitude of the modes must be around $10^{-5}$ when they exit the Hubble radius. Now the subHubble fluctuations (whether thermal or quantum) typically depend on the dimensionless quantity $\rho/M_p^4$. For instance, in traditional inflation,
\be \label{subhubble}
\cP_{\xi}\sim {\rho/M_p^4}\ .
\ee
Since one is required to produce a pretty large amplitude, this forces the scale of energy density to be reasonably close to the Planck scale. The typical scale of inflation is, for instance, $\cO(10^{-3})M_p$. The actual physical wavelength of the perturbations at this time, is therefore, quite short. Taking the example of inflation, the physical wavelength is given by
\be
l_{\mt{phys}}\sim H^{-1}\sim {M_p\over \sqrt{\rho}}\sim 10^6 l_p
\ee
where $l_p$ is the Planck length. Now suppose, just after such a mode-exit, the universe enters a radiation dominated expansion. In going down from a temperature of $\cO(10^{-3})M_p$ to the current temperature $\sim \cO(10^{-5})$ eV, the wavelength would approximately have stretched to a size of $\sim 10^{35} l_p $ which is still woefully short of the cosmological wavelengths we observe in sky today, in fact, approximately by a factor of 60 efoldings. This is obviously the precise reason why we need approximately that many inflationary efoldings after the modes that we see in the sky today exits the Hubble radius during inflation.

The crucial point is that very similar reasoning also applies to most other non-inflationary mechanisms of generating fluctuations, even if it is generated during contraction. In other words, at the moment the fluctuations exit the Hubble radius and freeze, typically their physical wavelengths are very small.
For instance, if (\ref{subhubble}) still applies (which by the way is the case for most scenarios as long as there is a single fluid/scalar field), for a symmetric bounce evolution, we would fall short exactly by 60 efoldings!
This indeed holds in our case as well. If we consider for example $M_s=10^{-3}M_p$ and the temperature range $10^{-4}<T/M_p<1$, we span the scales $10^3 l_p <l_{\mt{phys}}<10^6 l_p$.
Thus we, like most other bouncing/cyclic models, need to generate this asymmetry between expansion and contraction.

This could be arranged if the universe underwent a phase of inflationary expansion after the bounce. The advantage is that the inflationary epoch need not be near exponential, a power-law growth should be sufficient.  Another possibility is to have several phases of asymmetric cyclic growth as in {\it cyclic-inflation}~\cite{cyclic-inflation}.  For the cyclic inflation case, if the entropy production in a single cycle is large, then potentially all the CMB modes  approximately spanning  3 orders of magnitude ($\ell=2$ to $\ell\sim 2500$) can exit within the same cycle, and all our previous analysis then remains valid. If on the other had, the entropy production is small, then essentially the modes will exit approximately at the same temperature but in subsequent cycles. In this case, the spectral tilts will depend on how the turnaround energy scale changes in subsequent cycles. This is typically governed by underlying scalar field evolution~\cite{cyclic-inflation} and the spectral-tilt calculations will need to be revisited.

To summarize, we have considered a very simple thermal universe motivated by stringy physics. It provides a near scale invariant matter spectrum which could be red-tilted, and predicts a blue tilted gravity wave spectrum. Additionally, the expected range of the Hagedorn phase (in orders of magnitude) is tantalizingly close to what we are currently able to access in the CMB sky leading to the possibility of being able to see the ``edge effects'': a power enhancement at low multipoles as the Hagedorn phase gives way to more traditional growth of pressure with temperature, and  a running of the gravitational spectral tilt at high multipoles  as imprints of a quantum gravitational bounce!
\vs
{\bf Acknowledgments:}
This work was supported by the U.S. DOE Grant Nos. DE-FG02-87ER40328 and DOE/DE-FG02-94ER40823, the FPA 2005-02327 project (DGICYT, Spain), and the CAM/UCM 910309 project. AM is supported by the Lancaster-Manchester-Sheffield Consortium for Fundamental Physics under STFC grant ST/J000418/1.

\end{document}